\documentclass[a4paper]{spie}  

\usepackage{calc}
\usepackage{amsmath,amsfonts,amssymb}
\usepackage{graphicx}
\usepackage[colorlinks=true, allcolors=blue]{hyperref}

\title{Absolute and relative surface profile interferometry using multiple frequency-scanned lasers}

\author[$\bullet$]{Marek Peca}
\author[$\circ$]{Pavel Psota}
\author[$\circ$]{Petr Vojt\'i\v{s}ek}
\author[$\circ$]{V\'it L\'edl}

\affil[$\bullet$]{{\sc Eltvor}, T\'abor, Czech Republic}
\affil[$\circ$]{TOPTEC, Institute of Plasma Physics, Academy of~Sciences of the~Czech Republic}

\authorinfo{Corresponding author: Marek Peca, $\langle${\tt{mp@eltvor.cz}}$\rangle$}

\begin{document} 
\maketitle

\begin{abstract}
An interferometer has been used to measure the surface profile of generic object. Frequency scanning interferometry has been employed to provide unambiguous phase readings, to suppress etalon fringes, and to supersede phase-shifting. The frequency scan has been performed in three narrow wavelength bands, each generated by a temperature tuned laser diode. It is shown, that for certain portions of measured object, it was possible to get absolute phase measurement, counting all wave periods from the point of zero path difference, yielding precision of 2.7nm RMS over 11.75mm total path difference. For the other areas where steep slopes were present in object geometry, a relative measurement is still possible, at measured surface roughness comparable to that of machining process (the same 2.7nm RMS). It is concluded, that areas containing steep slopes exhibit systematic error, attributed to a combined factors of dispersion and retrace error.
\end{abstract}


\section{Objective}

The objective of present work was to measure the surface profile of a generic part using interferometer. There was a strong interest to perform an absolute interferometry, i.e. to resolve the exact optical path difference (OPD), without ambiguity of $n2\pi$ (integer number of skipped wavelengths). One of the considerable benefits of absolute measurement is the ability to resolve a complicated shape with sharp discontinuities, what in general is not possible by phase unwrapping\cite{malacara2007}.

Our intent was to perform absolute surface mapping, using the simplest possible setup, and with uncertainty comparable to high-grade optical parts ($\ll\lambda/100$). The aim was to reach results comparable or better than white-light interferometers\cite{graff2013wli,zygo2015wli}, without the need of mechanical scanning. Also, to provide at least comparable results with other absolute interferometers\cite{ptb2009multi,ptb2014lep,cern2004absint,xiaoli1998absint,yang2005absint}, either using a substantially simpler hardware, or at least providing a matrix of pixels (surface map), where other teams perform a scalar ranging only.

\section{Experiment and method}

\subsection{Measurement setup}

\begin{figure}[ht]\begin{center}\begin{tabular}{cc}
a)\raisebox{10pt-\height}{\includegraphics[width=100mm]{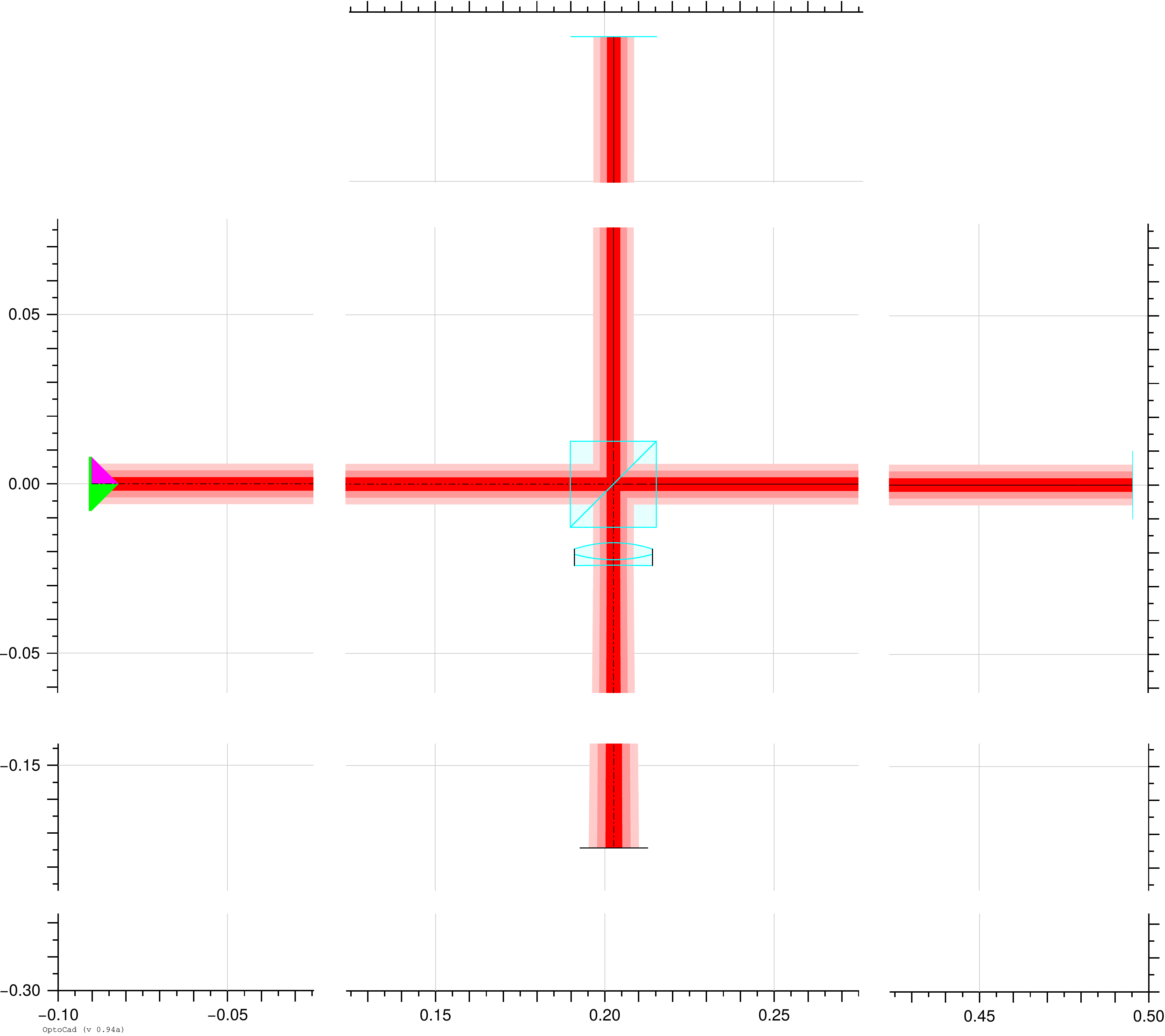}}&
b)\raisebox{10pt-\height}{\includegraphics[width=50mm]{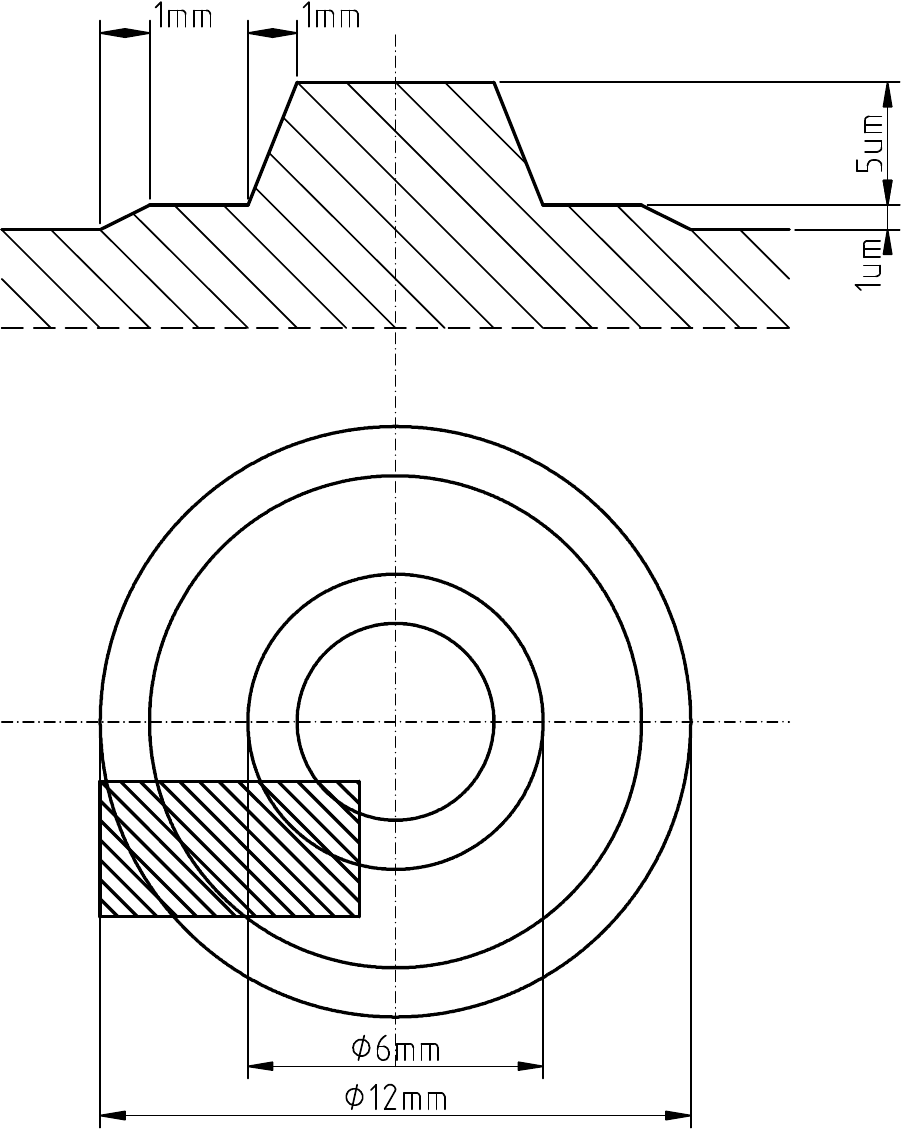}}
  \end{tabular}\end{center}
  \caption{\label{fig:1sch}
    (a) Interferometer layout; (b) drawing of measured object.}
\end{figure}

An object to be measured has to be reflective, and in approximate shape of a plane. The object is to be placed as a reflector in one arm of Michelson interferometer, Fig.~\ref{fig:1sch}a. In the other arm a reference flat is used. After combining both beams behind a beam splitter, an objective lens performs imaging from the object plane to the plane of image sensor (camera chip). The setup has been focused so that rays deflected by the object shall be focused onto the image sensor. The reference flat is somewhat displaced from zero OPD, thus making a clearly distinguishable absolute offset.

For the signal formation and processing we have chosen frequency-scanning interferometry (FSI\cite{malacara2007}). The reasons were mainly: (1.) easy and cheap temperature (or current) tuning of contemporary diode lasers; (2.) ability to replace phase shifters by scanning frequency only; (3.) ability to resolve $n2\pi$ ambiguity, and also we presumed (4.) the method will be more resistant to etalon fringes and superimposed false interference. Our expectation of~(4.)~is~based on fact, that spurious fringes resulting from different OPDs in system will generate in turn different fringe density in FSI, which shall be close to orthogonal to the desired OPD for wide frequency scans.

As a tunable laser source, 3 DFB diode lasers have been employed, $\sim2\,{\rm MHz}$ Lorentzian linewidth, $\Delta\lambda\approx1.5\,{\rm nm}$ of temperature tuning range. The nominal center frequencies of these DFB lasers were: $773\,{\rm nm}$, $785\,{\rm nm}$, $852\,{\rm nm}$. The DFB lasers were operated at constant current, variable temperature control. All of them were coupled and multiplexed to a single-mode fiber, illuminating the interferometer through a beam collimator. A part of the fiber light has been tapped off and led to a wavemeter in order to provide accurate actual frequency measurements. The declared accuracy of the wavelenght measurement is $5\times10^{-7}$. The camera is based on a 12-bit, APS image sensor, $6\times6\mu m$ pixels.

All the optical components were off-the-shelf items: beam splitter cube made of $25.4\,{\rm mm}$ N-BK7, and a $f=75\,{\rm mm}$ achromatic doublet made of N-BAF10, N-SF6HT. The system has been set up on an air-suspended breadboard table, and covered under a cardboard box as a rudimentary means of temperature and air flow control.

\subsection{Signal processing}

\begin{figure}[ht]\begin{center}\begin{tabular}{c}
      \includegraphics[scale=1.0]{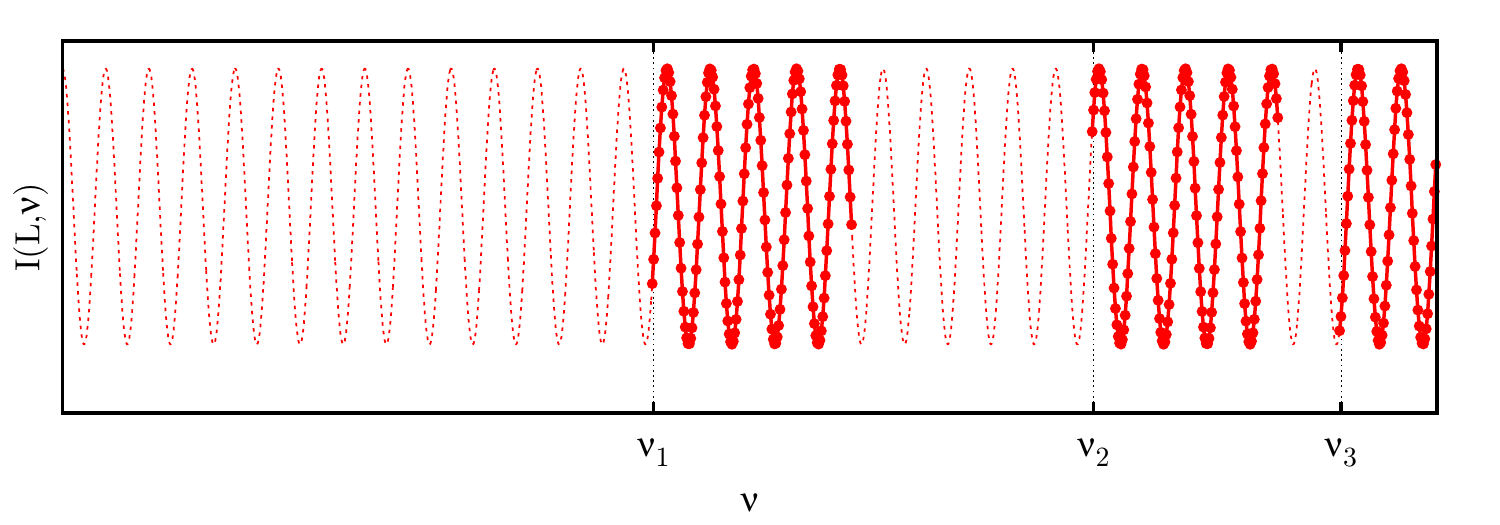}
  \end{tabular}\end{center}
  \caption[rms]{\label{fig:sigproc} 
    Illustration of FSI data sampled at distinct frequency bands.}
\end{figure}

At each pixel, a dependence of intensity on laser frequency is sampled at several discrete points, see idealized and exaggerated plot in Fig.~\ref{fig:sigproc}. The intensity function of ideal interferometer, having exactly one path difference (two arms, no spurious reflections), uniform power, no instrumental noise etc., is:

\begin{equation}
  I(L,\nu)\propto\cos{2\pi\frac{L}{\lambda}}+I_{DC}=\cos{\frac{2\pi}{c}L\nu}+I_{DC}.
\end{equation}

In our task $\nu$ is known variable and the OPD $L$ is a parameter to be estimated. The intensity $I(L)$ as a function of OPD is a cosine waveform with zero phase at zero OPD, as well as at (hypothetical) zero laser frequency. The ultimate goal of absolute FSI is to reconstruct the cosine from sparse measured samples -- ideally the waveform shall be extrapolated to origin ($\nu=0$) with phase error not exceeding $\pi$. Otherwise the integer ambiguity $n2\pi$ can hardly be detected from the data. The $L$ effectively plays a role of fringe repetition rate during the $\nu$ sweep.

Within our setup, we have few (three) narrow $\nu$ bands, separated by comparatively large gaps. Let us suppose we wish to extrapolate the waveform between two distant bands centered around $\nu_{1,2}$. We need to have an initial fringe rate estimate $\hat{L}_0$, and estimated fringe phase at the two exact frequencies: $\hat{\varphi}_{1,2}\approx\varphi_{1,2}=(2\pi/c)L\nu_{1,2}$. We must take into account the errors of all three prior estimates: $\hat{L}_0=L+\varepsilon_L$, $\hat{\varphi}_{1,2}=\varphi_{1,2}+\varepsilon_{\varphi_{1,2}}$. The linear extrapolation of phase can be only successful when the errors combined do not exceed $\pm\pi$:

\begin{equation}
\frac{2\pi}{c}|\varepsilon_L||\nu_2-\nu_1|+|\varepsilon_{\varphi_1}|+|\varepsilon_{\varphi_2}|<\pi.
\end{equation}

In practice $\varepsilon_{\varphi_{1,2}}$ will be negligible compared to $\varepsilon_L$, therefore the prerequisite for extrapolation over $\nu_1\dots\nu_2$ gap is:

\begin{equation}\label{eq:errpre}
  |\varepsilon_L|<\frac{c}{2|\nu_2-\nu_1|}.
\end{equation}

After closing the gap (without $n2\pi$ slip), we can throw away the original OPD estimate $\hat{L}_0$, and calculate the new one:

\begin{equation}
  \frac{2\pi}{c}\hat{L}_{12}|\nu_2-\nu_1|=\hat{\varphi}_2-\hat{\varphi}_1.
\end{equation}

The error bound of the new estimate $\varepsilon_L'=\hat{L}_{12}-L$ is:

\begin{equation}\label{eq:errpost} |\varepsilon_L'|=\frac{|\varepsilon_{\varphi_1}|+|\varepsilon_{\varphi_2}|}{\pi}\frac{c}{2|\nu_2-\nu_1|}.
\end{equation}

In any meaningful real-life setup $|\varepsilon_{\varphi_{1,2}}|\ll2\pi$, therefore the OPD estimate $\hat{L}$ is improved after the extrapolation (compare with Eq.~\ref{eq:errpre}), in proportion to phase errors. These extrapolation steps may be iterated more than once through frequency bands, ultimately finishing with a desired step to the known $\varphi(\nu=0)=0$ boundary condition, so the absolute disambiguation is closed.

\section{Surface reconstruction results}

\subsection{Object and measurement}

The object under test was a rotation symetrical part machined from EN~AW2030 aluminium by Nanotech~350FG CNC, according to a drawing in Fig.~\ref{fig:1sch}b. The object has been positioned in the interferometer to be approximately parallel with an image of reference flat, and with an absolute OPD of approximately $1\,{\rm cm}$. The main reason for such a large OPD compared to object depth range was twofold: (1.) to provide sufficiently high number of fringe periods over each $1.5\,{\rm nm}$ scan to improve orthogonality against etalon and false fringes; (2.) to stretch the goal of absolute interferometry (precision vs. total range).

\begin{figure}[ht]\begin{center}\begin{tabular}{cccc}
a) &\raisebox{10pt-\height}{\includegraphics[scale=0.5]{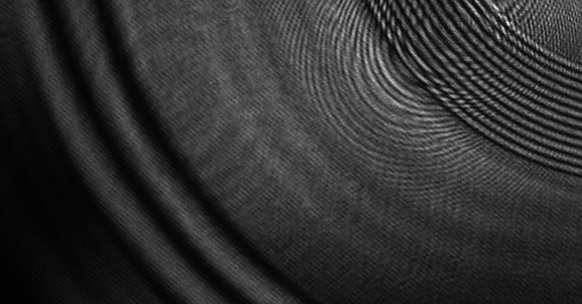}}\\
b) &\raisebox{15pt-\height}{\includegraphics[scale=1.0]{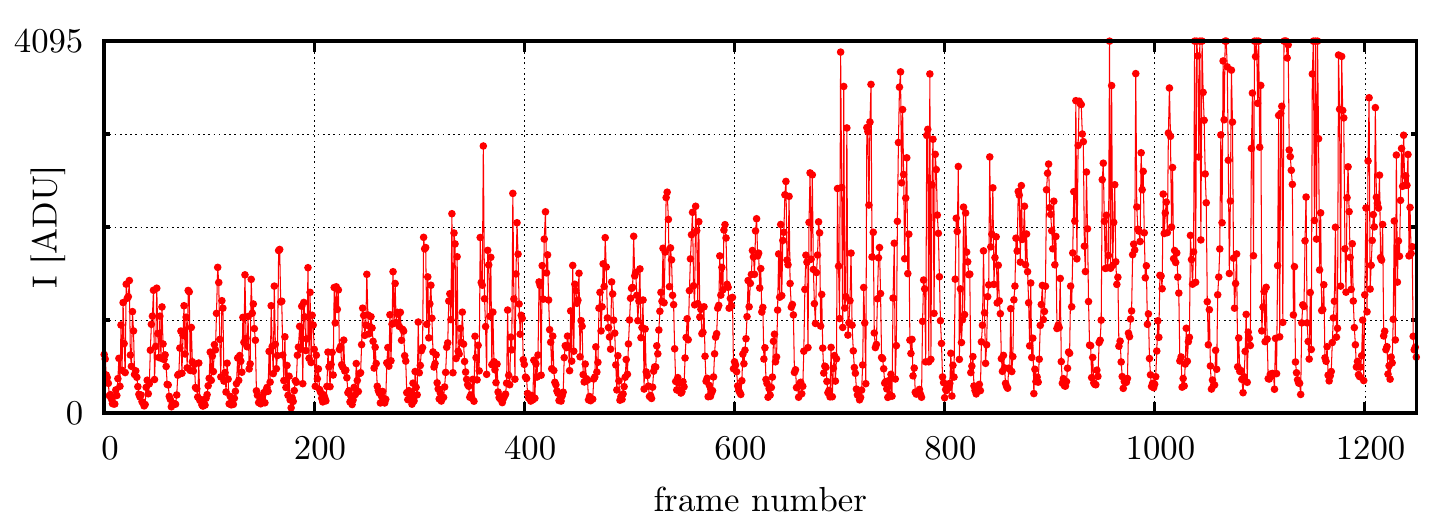}}
    \end{tabular}\end{center}
  \caption{\label{fig:03img}
    (a)~Sample image sensor frame, captured at $\lambda=852\,{\rm nm}$; (b)~evolution of pixel intensity vs. wavelength scan}
\end{figure}

Each of the 3 DFBs has been swept in frequency in the whole range of $\approx1.5\,{\rm nm}$ in 1250 approximately equidistant steps. At each step (frequency point) one camera frame has been captured. A cropped area of $582\times304$ pixels was selected for analysis (hatched rectangle in Fig.~\ref{fig:1sch}b, Fig.~\ref{fig:03img}a). There were several imperfections in the data observed. First, strong etalon fringes were visible in great part of the frame, often spanning over tens of \% of the signal amplitude (Fig.~\ref{fig:03img}a). Also, a saturation occured in considerable number of the data points. Second, due to a data acquisition problems there were large portions of reference wavelength measurements missing, so they were interpolated by a 2nd-order polynomial fit. An example of one 1250 samples long frequency scan is in Fig.~\ref{fig:03img}b. It displays an intensity of one selected camera pixel over the 1250 different DFB temperatures (frequencies).

\subsection{Absolute interferometry}

In the absolute FSI, each pixel plays on its own, without any interpolation or phase unwrapping with neighbour pixels. In effect, the absolute FSI is physically and mathematically equivalent to CW chirp lidar (as well as radar), just the sweep time, span and absolute range may differ several orders of magnitude (mm vs. km).

\begin{figure}[ht]\begin{center}\begin{tabular}{cccc}
a)&\raisebox{10pt-\height}{\includegraphics[width=75mm]{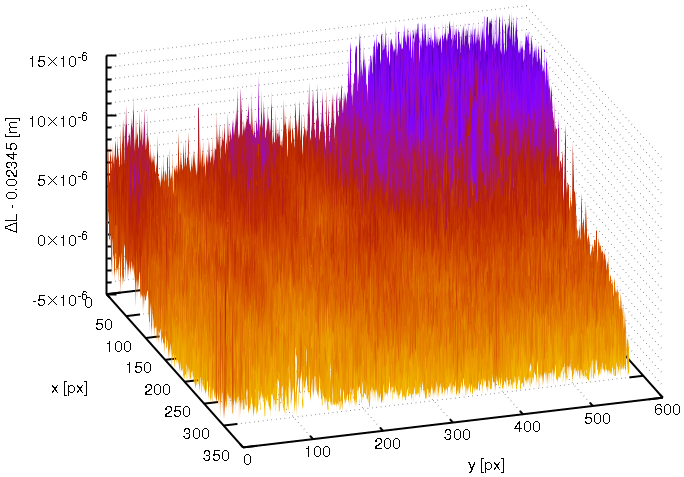}}&
d)&\raisebox{10pt-\height}{\includegraphics[width=75mm]{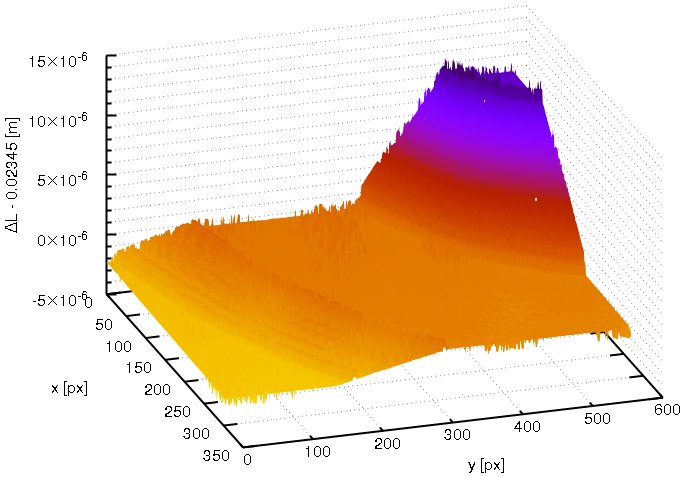}}\\
b)&\raisebox{10pt-\height}{\includegraphics[width=75mm]{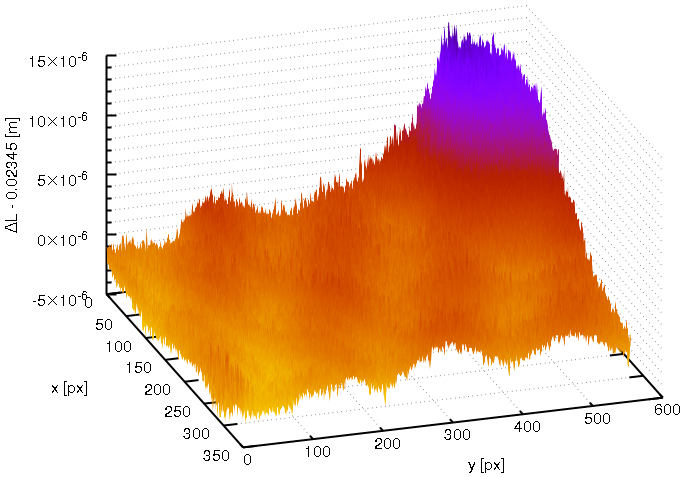}}&
e)&\raisebox{10pt-\height}{\includegraphics[width=75mm]{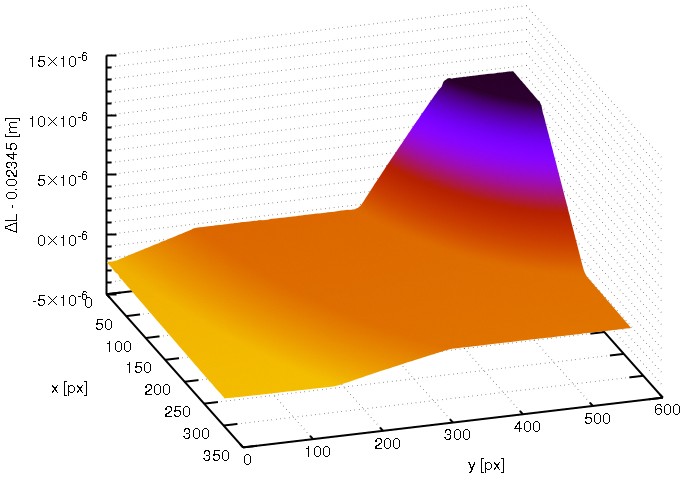}}\\
c)&\raisebox{10pt-\height}{\includegraphics[width=75mm]{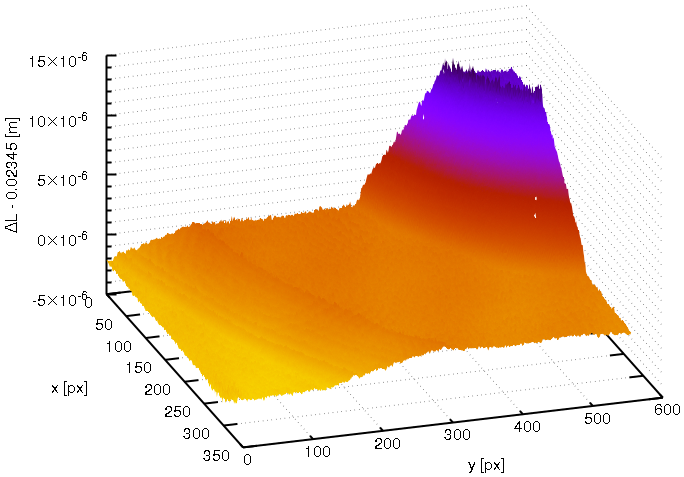}}&
f)&\raisebox{10pt-\height}{\includegraphics[width=75mm]{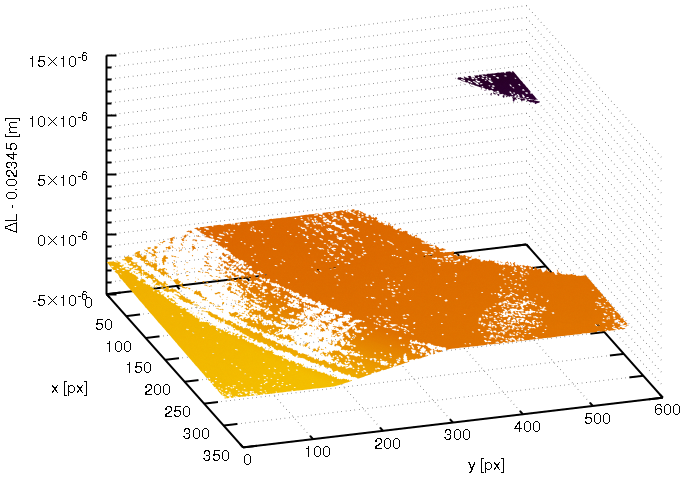}}\\
  \end{tabular}\end{center}
  \caption{\label{fig:z}
    (a$\dots$d)~Iterations of absolute interferometry; (e)~unwrapped relative interferometry; (f)~subset of absolute depth map after outlier removal.}
\end{figure}

Let us observe in the following figures, how did the surface map evolve during sequence of processing steps, beginning at the longest unambiguous synthetic wavelength, going down to the final wavelength of the actual light. Fig.~\ref{fig:z}a shows the first coarse estimate of OPD, resulting solely from a fringe density of the sweep of only one DFB. The OPD detected is $\approx23.5\,{\rm mm}$, and a mean square error (compared by subtracting later refined shape) is $\sigma_0=4\,{\rm \mu m}_{RMS}$. Fig.~\ref{fig:z}b displays a refinement of surface by subsequent iteration, where $773\,{\rm nm}$ sweep has been unambiguously joined with $785\,{\rm nm}$ sweep, $\sigma_1=1\,{\rm \mu m}_{RMS}$. Fig.~\ref{fig:z}c shows another iteration, where all three wavelength sweeps have been joined together, and their fringe phase has been unambiguously joined from $852\,{\rm nm}$ through $773\,{\rm nm}$, resulting deviation $\sigma_2=0.43\,{\rm \mu m}_{RMS}\approx 0.56\,\lambda$. At this point it is evident, that unambiguous {\em absolute} measurement, allowing to extrapolate fringe phase to zero frequency with an error less than $\pm\pi$, is not possible for every frame pixel.

\begin{figure}[ht]\begin{center}\begin{tabular}{cccc}
a) &\raisebox{10pt-\height}{\includegraphics[scale=0.75]{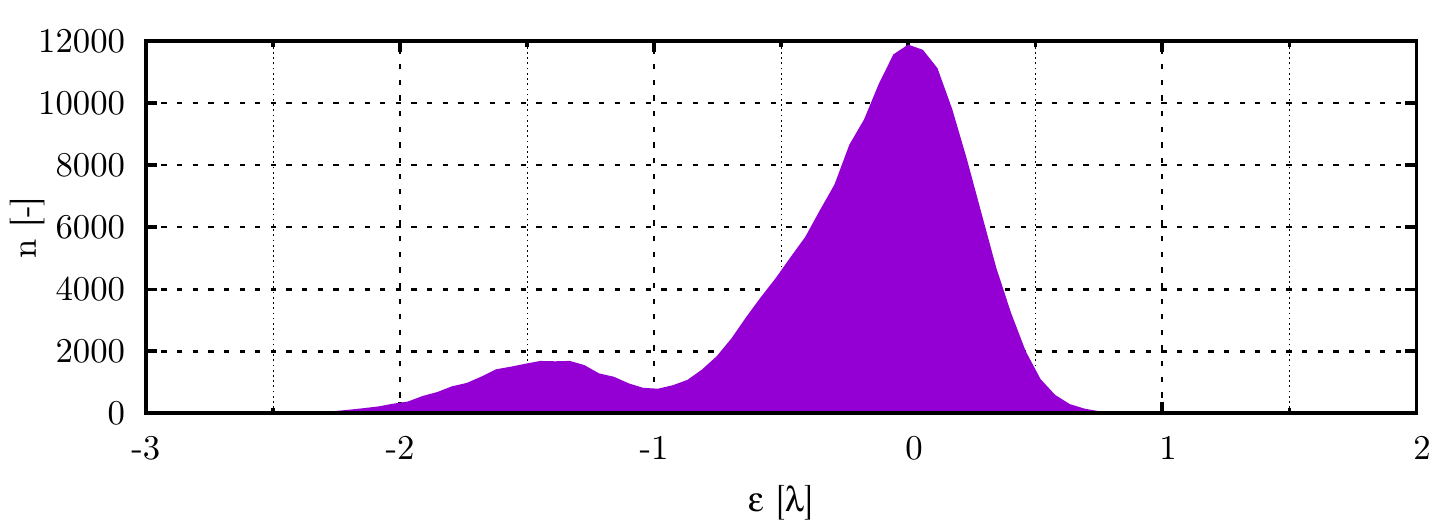}}\\
b) &\raisebox{10pt-\height}{\includegraphics[scale=0.75]{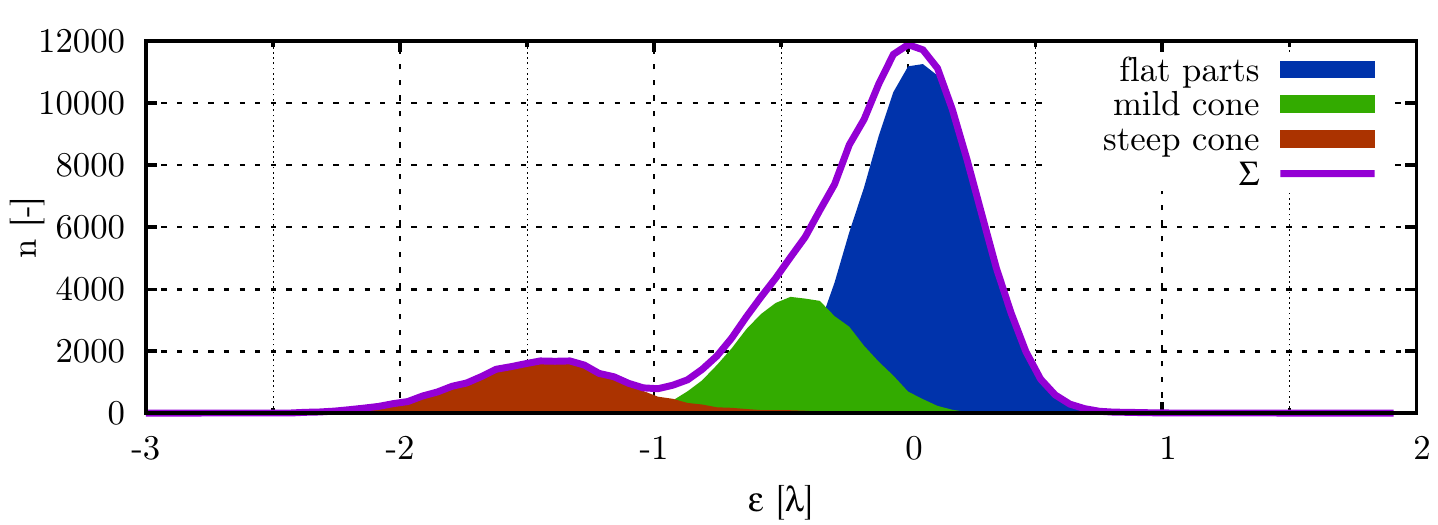}}
  \end{tabular}\end{center}
  \caption{\label{fig:10hist}
    Error of absolute OPD estimate: (a)~all pixels in one histogram; (b)~image subdivided into three areas: flat areas, steep conic slope, mild conic slope.}
\end{figure}

Fig.~\ref{fig:z}d shows the -- in large subset of pixels unsuccessful -- attempt to extrapolate the fringe phase to zero frequency, and to resolve the exact surface induced OPD without $n2\pi$ ambiguity. It has been observed, that there were frequent phase slips of $\pm2\pi$, and smaller number of $\pm2\times2\pi$, and a negligible number of larger slips. It is obvious, that while in the flat parallel areas the $n2\pi$ spikes are sparse and likely to be discovered by a suitable post-processing algorithm, the areas of both conical slopes contains a large amount of {\em systematic error}. A better insight into this fact is given by the histogram in Fig.~\ref{fig:10hist}. Let us search for the root cause of this failure by looking on the same data through the eyes of relative interferometry.

\subsection{Relative interferometry}

While our ultimate goal is to measure generic profiles in an absolute manner, and to provide a viable alternative to existing absolute techniques, such as white-light interferometry\cite{graff2013wli,zygo2015wli}, the actual object in our experiment had a smooth profile thus allowing to perform relative phase unwrapping, and so to reconstruct a smooth OPD surface.

Although the FSI has not been equipped with phase shifters, a relative phase of each of the DFB scans alone can be used as a robust and well averaged phase for subsequent phase unwrapping. Since our object was smooth enough, there was no need for sophisticated robust 2D unwrapping algorithms (such as residue method\cite{malacara2007}). We have just performed two 1D unwrappings along the cartesian axes one after another. Each of the three DFB scans thus produced a {\em seemingly} well-looking surface map, e.g. Fig.~\ref{fig:z}e. Soon it has been revealed, that this shape is again incorrect, due to a retrace error\cite{kreischer2013retrace,liu2009retrace}.

We have compared the surface map obtained from a DFB scan, converted to range in meters $L=\varphi\lambda/(2\pi)$, with the former absolute map of Fig.~\ref{fig:z}d (after subtraction of common offset, caused by the negligence of the relative approach). For the sake of clarity, the areas where mutual error was smaller than $\lambda/2$ have been selected, and plotted in Fig.~\ref{fig:z}f. It is confirmed the flat parallel areas mostly agree, while the slopes do not, the steeper being worse.

At this point it is necessary to distinguish between two different kinds of outcome. It depends, on what we exactly claim to measure: (a) depth map of OPD;
(b) real object's surface depth.

Due to the retrace error, there is a difference between (a) and (b), unless the object feature fulfill some special conditions, like in our case being parallel to projected reference flat. Moreover it has been observed, that there are observable {\em differences between the OPD maps among different wavelengths}. The only possible explanation for this is refractive index dispersion within the glass elements (beam splitter, lens doublet), acting in conjunction with the retrace error. The subject is examined in more detail in Sec.~\ref{sec:retr}.

\begin{figure}[ht]\begin{center}\begin{tabular}{c}
      \includegraphics[width=166mm]{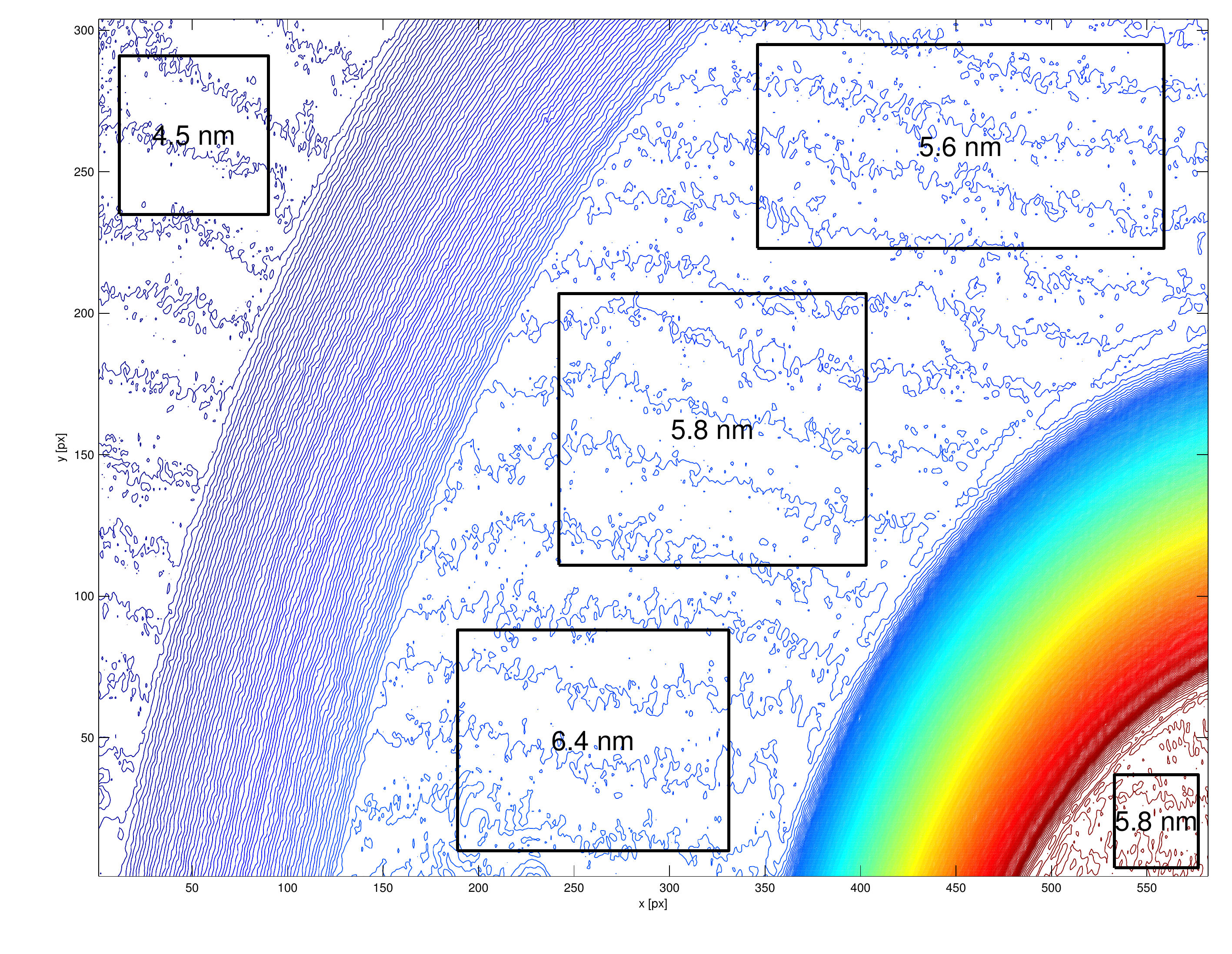}
    \end{tabular}\end{center}
  \caption[rms]{\label{fig:13rms} 
  OPD depth map retrieved from relative phase measurements. In rectangular areas, residual RMS roughness is displayed. Contour spacing is $30\,{\rm nm}$. {\em All values designate OPD -- object features correspond to $1/2$ of OPD.}}
\end{figure}

In order to evaluate the OPD measurement precision regardless of the retrace error, we have averaged the three OPD maps together, plotted in Fig.~\ref{fig:13rms}. Within each of the delimited rectangle areas, an ideal plane has been fit and subtracted, and the residual surface roughness has been calculated (without any other spaial filtering). Near $x=350, y=180$ there can be seen a very slight remnant of spurious spherical fringes. It shall be noted, that seemingly severe etalon fringes near $x=550, y=50$ area have had no visible impact in the result. The average performance in OPD, divided by two to correspond to a surface rougness, is $\sigma=2.7\,{\rm nm}_{RMS}$. This figure is already comparable to the precision of particular aluminium machining process.

\section{Retrace error and dispersion}\label{sec:retr}

\subsection{Observable error}

\begin{figure}[ht]\begin{center}\begin{tabular}{c}
      \includegraphics[width=120mm]{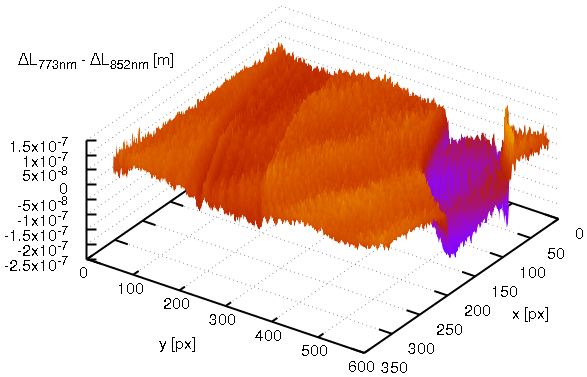}
    \end{tabular}\end{center}
  \caption[rms]{\label{fig:14dmap}
    Difference between OPD measured at $\lambda_1=773\,{\rm nm}$ minus OPD at $\lambda_2=852\,{\rm nm}$}
\end{figure}

\begin{figure}[ht]\begin{center}\begin{tabular}{cccc}
      a)\raisebox{10pt-\height}{\includegraphics[scale=0.54]{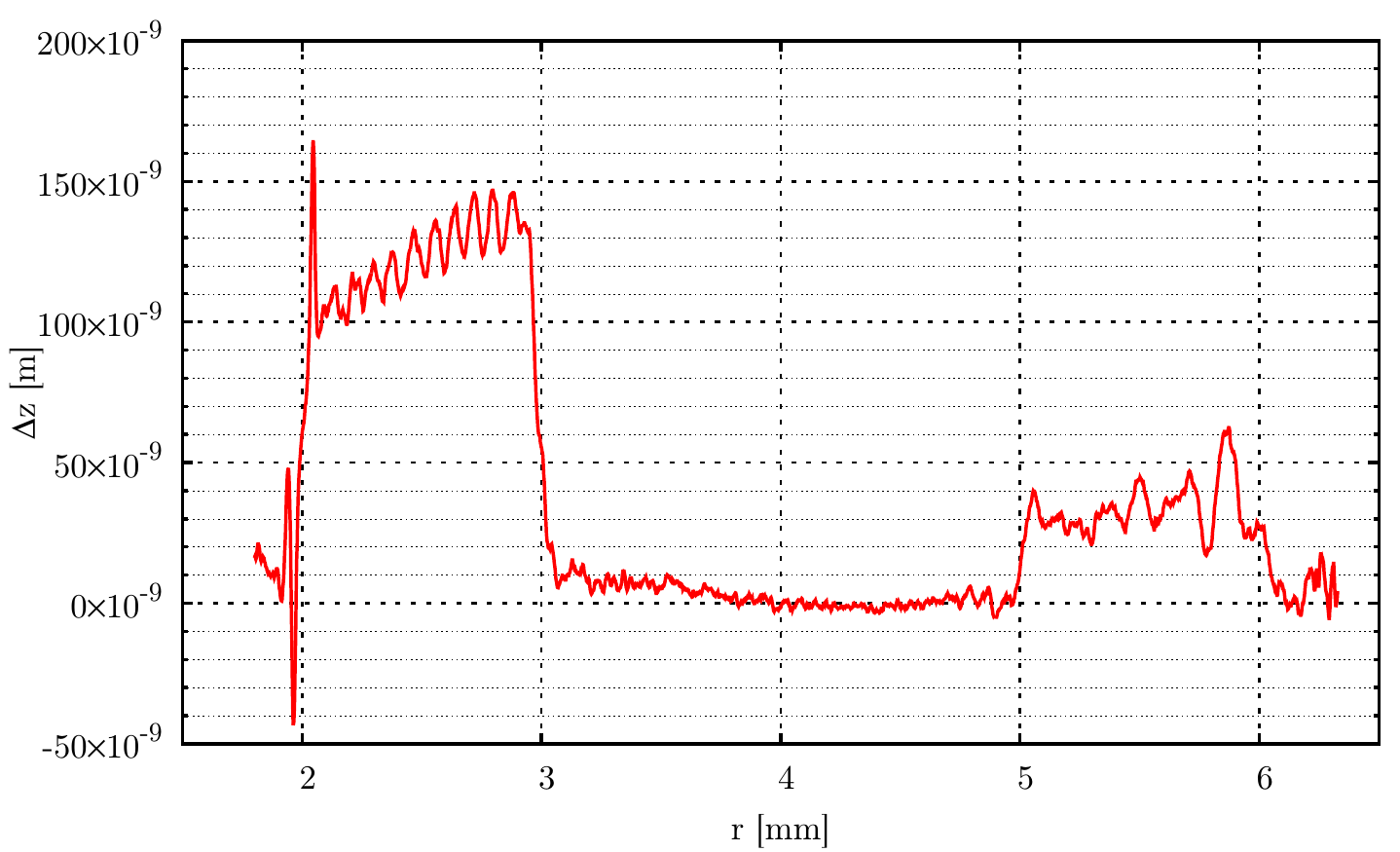}}&
      b)\raisebox{10pt-\height}{\includegraphics[scale=0.54]{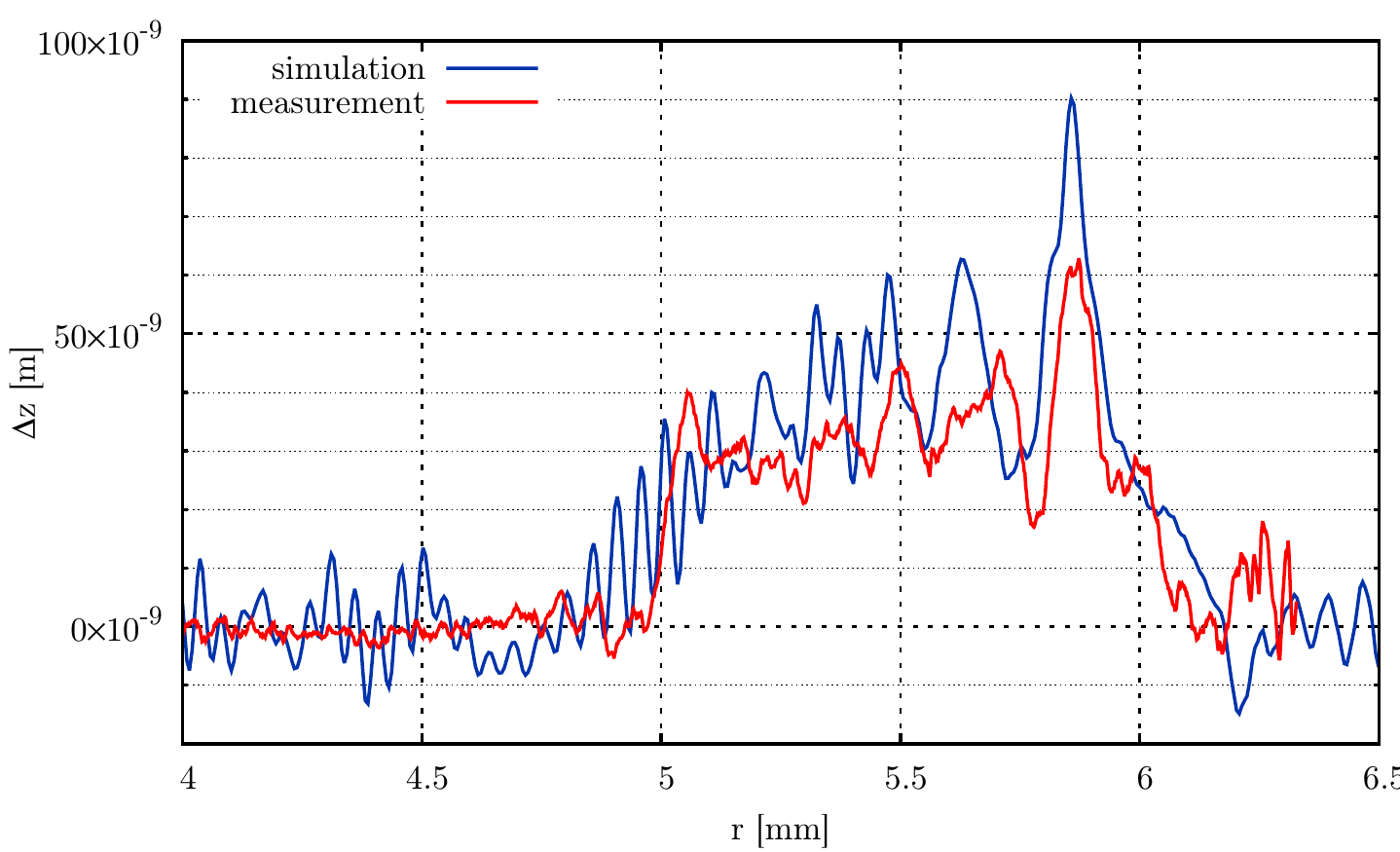}}
    \end{tabular}\end{center}
  \caption{\label{fig:15disp}
    Difference between OPD measured at $\lambda_1=773\,{\rm nm}$ minus OPD at $\lambda_2=852\,{\rm nm}$ -- radial projection; (a)~measurement; (b)~measurement vs. simulation.}
\end{figure}

The systematic error observed in absolute depth data led us to examine, how do compare the relative depth maps, compared among the distinct wavelengths. For the purpose, each of the DFB sweeps is regarded as a ``single-wavelength'' measurement, thanks to its relatively narrow scan bandwidth. In the following, the relative OPD maps at two distant wavelengths ($773\,{\rm nm}, 852\,{\rm nm}$ will be compared. Fig.~\ref{fig:14dmap} shows the two OPDs converted to length and subtracted (after ad hoc elimination of unknown absolute term). A projection of depth differences to a radial plane is plotted in Fig.~\ref{fig:15disp}.

It can be clearly seen, that both slope aread exhibit a consistent systematic error, while the flat areas do correspond among the different wavelengths. The steeper conical slope of $5\,{\rm \mu m}/1\,{\rm mm}$ causes a difference of $\approx113\,{\rm nm}$, the other slope of $1\,{\rm \mu m}/1\,{\rm mm}$ causes a difference of $\approx36\,{\rm nm}$ in OPD. The observed difference is caused by different retrace path at different wavelengths due to material dispersion. We are convinced that the only way to compensate for this kind of errors is to take the optical design of interferometer into account. The shape of the {\em OPD is a function of wavelength}.

\subsection{Simulation}

In order to verify the mangitude of dispersion effect on retrace error, the measurement setup has been modelled and propagation of laser beams simulated. The exact dimensions of interferometer have not been preserved, so at least an approximate model has been captured. The distance between camera, lens and object plane has been determined from known magnification, the least certain point was the distance between the lens and beam splitter. The setup has been tweaked a bit to get approximately focused beams originating at object plane.

The model has been entered as a script into a tiny open-source, Fortran-95 tool OptoCad\cite{schilling2002optocad}, Fig.~\ref{fig:1sch}a. The OptoCad is a geometrical optic tool only: it is able to propagate Gaussian beams and ideal rays. First, we have tried to get a confirmation of the observed difference in OPD for the two wavelengths, by transmission of artificially slanted ray beams, but without any success: the resulting OPD differences were like order of magnitude higher than measured.

Therefore, we have diverted to an accompanying tool to OptoCad, called WaveProp\cite{hild2008waveprop}, what is a Fourier optics tool. The WaveProp is again a tiny scripted tool, which works in paraxial approximation (the wave propagation vector is assumed to be always parallel to axis $z$, even for oblique beams). All calculations are performed in 2D, and with heavy usage of FFT. Our simulation has been limited by computer memory size to $4096\times4096$ complex amplitude points. The steeper of the two slopes has been difficult to simulate due to aliasing and probably also interpolation effects. Thus, we have entered a simplified model of object, containing only the $1\,{\rm \mu m}/1\,{\rm mm}$ conical slope.

\begin{figure}[ht]\begin{center}\begin{tabular}{ccc}
      a) \raisebox{10pt-\height}{\includegraphics[height=46mm]{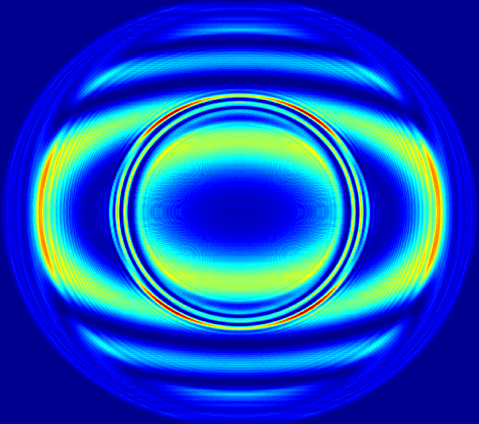}} &
      b) \raisebox{10pt-\height}{\includegraphics[height=46mm]{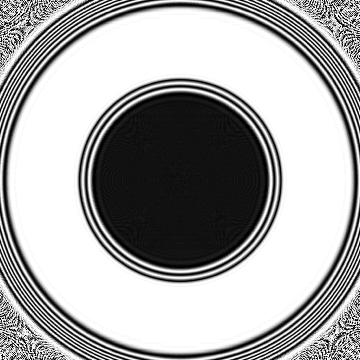}} &
      c) \raisebox{10pt-\height}{\includegraphics[height=46mm]{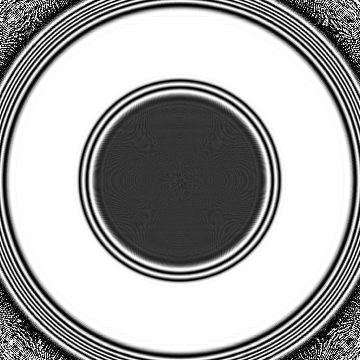}}
    \end{tabular}\end{center}
  \caption[rms]{\label{fig:16sim} 
  Simulated interference fringes; (a)~generic image-plane image; (b)~normalized fringes at $\lambda=773\,{\rm nm}$, (b)~at $\lambda=852\,{\rm nm}$.}
\end{figure}

In Fig.~\ref{fig:16sim}a there is an example, how does the interference pattern look like in image plane, when a Gaussian beam illuminates the interferometer, and a slanted mirror is in the reference arm. The beam has been then replaced by an ideal, aperture limited planar wavefront $\Psi(x,y)=1$. The object has been modelled as an ideal phase displacement, neglecting obliquity:

$$\Psi'(x,y)=\Psi(x,y)\exp{i\Delta\xi},\qquad \Delta\xi=4\pi\frac{z_{obj}}{\lambda}.$$

In the image plane, the two split beams were compared in phase, and the amplitude of interfering fields has been ignored:

$$\Delta\varphi={\rm arg}\left[\Psi_1(x,y)\Psi_2^*(x,y)\right].$$

The Fig.~\ref{fig:16sim}bc shows resulting normalized fringes $I=cos(\Delta\varphi)$ for the two different wavelengths. The $\Delta\varphi$ has been taken as the simulated interferometric readout. The same procedure as with the real life data has been performed with the simulated $\Delta\varphi$: OPD depth map has been unwrapped, converted to length units and the two wavelength's results have been subtracted. The result, converted and averaged to a radial plane, has been plotted together with the corresponding measured data in Fig.~\ref{fig:15disp}b. The agreement between simulated and real OPD differences is good, especially noting the maximum value is mere $36\,{\rm nm}$, i.e. $\lambda/20$.

\section{Conclusion}

\subsection{Results}

The absolute OPD measurement using our FSI has been achieved in most of the flat portions of the object under test, resolving unambiguously the total range of $2\times11.75\,{\rm mm}$ with the precision of $\sigma=2\times2.7\,{\rm nm}_{RMS}$. In other words, the total range of $3\times10^4\lambda$ has been resolved down to $<\lambda/2$, and a subsequent phase interpolation yielded the final precision of $\lambda/280$. The result is comparable to other similar FSI\cite{ptb2009multi}, as well as to different methods\cite{graff2013wli}, while our setup is comparatively simpler. It would be interesting to compare absolute and long-term performance with other non-imaging, scalar, but very accurate methods\cite{ptb2014lep,xiaoli1998absint,yang2005absint}. Current data, experimental conditions and object uncertainty do not allow immediate comparison, further experiments would be necessary.

The seemingly severe etalon fringes and pixel intensity saturation were suppressed surprisingly well during our FSI processing, leaving artefacts comparable in amplitude to a random looking noise.

The absolute FSI has yielded wrong results in the steep slope areas due to the retrace error. The relative approach to phase unwrapping was still possible thanks to the smooth object, however the outcome OPD does not correspond exactly to the surface profile due to the retrace error.

The precision $\sigma=2.7\,{\rm nm}_{RMS}$ of both methods is comparable to known machining roughness, so in order to characterize micro-roughness of the method, a well polished mirror shall be used instead. At this level of accuracy, the thermal expansion of structure and refractive index of air tend to enter the game, so before the assessment of absolute and long-term stability, a well controlled environment shall be provided.

Finally, the effect of different retrace error among different wavelength has been observed, and later simulated. The agreement of measured and simulated dispersion of the retrace looks promising, and we anticipate possible usage of simulated interferometer model in further processing.

\subsection{Further steps}

In future work, we would like to address the following enhancements:

\begin{itemize}
\item computational compensation of retrace error for both relative as well as absolute FSI;
\item characterization of micro-roughness (using a polished object);
\item long-term stability, repeatability (appropriate thermal and air control);
\item the same FSI with intentionally reduced laser coherence, in order to suppres even better the spurious fringes.
\end{itemize}

\acknowledgments

Special thanks go to Roman Dole\v{c}ek for machining the Fig.~\ref{fig:1sch}b part, and to Roland Schilling for providing his helpful and inspirative optical software\cite{schilling2002optocad,hild2008waveprop} as an open-source.

\bibliography{report} 
\bibliographystyle{spiebib} 

\end{document}